\documentclass[epj]{svjour}
\usepackage{graphics}
\newcommand{\be}{\begin{equation}}
\newcommand{\ee}{\end{equation}}
\newcommand{\bea}{\begin{eqnarray}}
\newcommand{\eea}{\end{eqnarray}}

\newcommand{\g}{\gamma}
\newcommand{\f}{\frac}

\newcommand\lr[1]{{\left({#1}\right)}}
\begin{document}
\title{Heavy-quark energy loss in pQCD and SYM plasmas}
\author{Cyrille Marquet
\thanks{C.M. is supported by the European Commission under the FP6 program,
contract No. MOIF-CT-2006-039860.}%
}                     
%
%
\institute{Institut de Physique Th\'eorique, CEA/Saclay, 91191 Gif-sur-Yvette cedex, France\\
Department of Physics, Columbia University, New York, NY 10027, USA\\
E-mail: cyrille@phys.columbia.edu}
\date{}
%
\abstract{
We consider heavy-quark energy loss and $p_\perp-$broadening in a strongly-coupled $N=4$ Super Yang Mills (SYM) plasma, and the problem of finite-extend matter is addressed. When expressed in terms of the appropriate saturation momentum, one finds identical parametric forms for the energy loss in pQCD and SYM theory, while $p_\perp-$broadening is radiation dominated in SYM theory and multiple scattering dominated in pQCD.
} 
\maketitle
\section{Introduction}
\label{intro}

It is often argued that hard probes such as heavy quarks are understood well enough
to provide clean measurements of the properties of the quark-gluon plasma formed in Au+Au collisions at RHIC. In particular, they could help determine whether the plasma is weakly
or strongly coupled. Results on bulk observables like the elliptic flow or the shear viscosity already prompted claims that the plasma is strongly coupled.

With hard probes, it is indeed unclear if the perturbative QCD (pQCD) approach
can describe the suppression of high$-p_\perp$ particles, in particular for heavy-quark production. High$-p_\perp$ electrons from charm and bottom mesons decays seem to indicate a similar suppression for light, charm and bottom quarks. By contrast in pQCD, the heavier
the quark the weaker the suppression.

This motivates to think about strongly coupled plasmas. The tools to address the strong coupling dynamics in QCD are quite limited, however for the $N=4$ Super-Yang-Mills (SYM) theory, the
AdS/CFT correspondence is a powerful approach. The findinds for the strongly-coupled SYM plasma may provide insight for gauge theories in general, and some aspects may even be universal, like the lower bound of the shear viscosity \cite{Kovtun:2004de}.

In this work we study the energy loss of a very energetic heavy quark propagating through
the SYM plasma. For comparison, the pQCD results are recalled in Section \ref{sec:1}.
The trailing-string picture of heavy-quark energy loss in a static infinite-extend plasma is introduced in Section \ref{sec:2}, and a partonic interpretation in terms of the saturation momentum is given which allows to infer the plasma length dependence of the energy loss when considering finite-extend matter.

\section{Heavy-quark energy loss in a weakly-coupled QCD plasma}
\label{sec:1}

In this section we consider a heavy quark of energy $E$ and mass $M,$ propagating
through a weakly-coupled QCD plasma and losing energy at the rate $-dE/dt$ due to
the interaction with the medium. We recall the pQCD results as well as the underlying
multiple scattering picture.

\subsection{The heavy quark wave function}

\begin{figure}
\begin{center}
\resizebox{0.7\columnwidth}{!}{\includegraphics{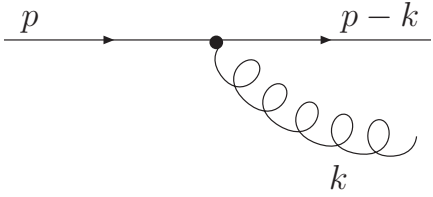}}
\caption{Virtual gluon fluctuation in the heavy quark wave function, with energy
$\omega$ and transverse momentum $k_\perp.$ The coherence time $\omega/k_\perp^2$
measures the virtuality of the fluctuation.}
\label{fig:1}
\end{center}
\end{figure}

Let us start with the heavy quark wave function in QCD. At lowest order in $\alpha_s,$
quantum fluctuations consists of a single gluon (see Fig.\ref{fig:1}),
whose energy we denote $\omega$ and transverse momentum $k_\perp.$ The virtuality of that fluctuation is measured by the coherence time, or lifetime, of the gluon 
\be
t_c=\omega/k_\perp^2\ .
\ee
Short-lived fluctuations are highly virtual while longer-lived fluctuations are more
easily put on shell when the heavy quark interacts. The probability of the fluctuation
is
\be
P\!=\!P(M\!=\!0)\lr{1\!+\!\f{\omega^2}{\g^2k_\perp^2}}^{-2}\mbox{with }
P(M\!=\!0)\!\sim\!\alpha_sN_c\ .
\ee
We have introduced the Lorentz factor of the heavy quark
\be
\g=E/M\ .
\ee
Compared to massless quarks, the fluctuations with $\omega>\g k_\perp$ are suppressed
in the wave function. This means that when gluons are put on-shell, they are not radiated
in a forward cone around the heavy quark. This suppression of the available phase space for radiation, the {\it dead-cone} effect, implies less energy loss for heavier quarks
\cite{Dokshitzer:2001zm}.

\subsection{Medium-induced gluon radiation}

In pQCD, medium induced gluon radiation is due to multiple scatterings of the virtual gluons
\cite{Baier:1996sk}. If, while undergoing multiple scattering, the virtual gluons pick up enough transverse momentum to be put on shell, they become emitted radiation. The accumulated transverse
momentum squared pickep up by a gluon of coherence time $t_c$ is
\be
p_\perp^2=\mu^2 \f{t_c}{l}\equiv\hat{q}\ t_c
\ee
where $\mu^2$ is the average transverse momentum squared pickep up in each scattering, and
$l$ is the mean free path. These medium properties are involved through the ratio
$\hat{q}=\mu^2/l,$ this is the only relevant information about the medium. In terms of the temperature $T,$ one has
\be
\hat{q}\sim\alpha_s T^3
\ee
and at RHIC temperature, the value is $\hat{q}\sim 1\ \mbox{GeV}^2/\mbox{fm}$ (however
$5-10\ \mbox{GeV}^2/\mbox{fm}$ seems favored by RHIC data).

Since only the fluctuations which pick up enough transverse momentum are freed ($k_\perp<p_\perp$),
the limiting value can be obtained by equating $k_\perp^2$ with
$p_\perp^2=\hat{q}\omega/k_\perp^2:$
\be
k_\perp<(\hat{q}\omega)^{1/4}\equiv Q_s(\omega)\ .
\ee
The picture is that highly virtual fluctuations with $k_\perp>Q_s$ do not have time to pick up enough $p_\perp$ to be freed, while the longer-lived ones with $k_\perp<Q_s$ do. That transverse momentum $Q_s$ which controls which gluons are freed and which are not is called the saturation scale.

\subsection{Heavy-quark energy loss}

When applying this picture to heavy quarks, one sees that due to the dead cone effect,
the maximum energy a radiated gluon can have is $\omega=\g k_\perp=\g Q_s,$ and its
coherence time is $t_c=\g/Q_s.$ This, combined with the probability $\alpha_s N_c$ to have the gluon fluctuation in the wave function, allows to estimate the heavy-quark energy loss:
\be
-\f{dE}{dt}\propto\alpha_sN_c\frac{\gamma Q_s}{\gamma/Q_s}=\alpha_s N_c Q_s^2\ .
\ee
The saturation momentum in this formula is that of the fluctuation which dominates the energy loss, and should be evaluated at $\omega=\g Q_s.$ Using $Q_s^2=\sqrt{\hat{q}\omega}$ one can express that value of the saturation scale in terms of $T$ and $\gamma$ only:
\be
Q_s=(\hat{q}\gamma)^{1/3}\ .
\ee
In the following, that particular value is what we mean by $Q_s.$ We shall also use $t_c$ to denote the coherence time of the dominant gluonic fluctuation:
\be
t_c=\g^{2/3}/\hat{q}^{1/3}\ .
\ee

\subsection{The case of finite-extend matter}

When the heavy quark is propagating through a plasma of finite length $L,$ the discussion above has to be modified \cite{Baier:1996kr}. If $L>t_c,$ the gluons which dominated the energy loss in the infinite matter case have the time to form before the heavy quark exits the plasma, and therefore the matter is effectively of infinite extend and the above results are unchanged. If $L<t_c,$ only shorter-lived fluctuations, with less energy $\omega<Lk_\perp^2$ can contribute to the energy loss. The dominant ones have a coherence time $\omega/k_\perp^2=L,$ and the transverse momentum squared they picked up is $\hat{q}L.$ This defines the saturation scale in the finite matter case:
\be
Q_s^2=\hat{q}L\ .
\ee
All gluons with $k_\perp<Q_s$ are freed and the maximum energy of a radiated gluon is
$\omega=Lk_\perp^2=LQ_s^2$ which gives
\be
-\f{dE}{dt}\propto\alpha_sN_c\frac{L Q_s^2}{L}=\alpha_s N_c Q_s^2\ .
\ee
This formula is the same as before except that the saturation scale is different. In the finite matter case, $Q_s$ does not depend on $E/M$ anymore, but on $L$ instead.

Note that the saturation momentum discussed here is not the traditional saturation momentum characterizing the small$-x$ part of a hadronic wave function. $\hat{q}L$ has been denoted $Q_s^2$
because when writting $\hat{q}L$ in terms of the gluon density per unit of transverse area in the plasma, one finds the same expression than when writting the saturation scale in terms of the gluon density per unit of transverse area in a hadron.

\subsection{Heavy-quark $p_\perp-$broadening}

From the above discussion, the radiative $p_\perp-$broadening of the heavy quark is easily
estimated:
\be
\f{dp_\perp^2}{dt}\propto \alpha_s N_c\f{dQ_s^2}{dt}=\alpha_s N_c\ \hat{q}
\ee
where $t=t_c$ in the infinite matter case and $t=L$ in the finite matter case. The result is the same in both cases because the broadening is due to multiple scatterings which give local
$p_\perp$ kicks to the fluctuations. However radiative $p_\perp-$broadening is irrelevant in pQCD
(we mention it for future comparisons) with $\alpha_s\ll1,$ because what is dominant is
$p_\perp-$broadening due to multiple scattering of the heavy quark itself:
\be
\f{dp_\perp^2}{dt}\propto\f{dQ_s^2}{dt}=\hat{q}\ .
\ee

\section{Heavy-quark energy loss in a strongly-coupled SYM plasma}
\label{sec:2}

In this section we compute the rate of energy loss of the heavy quark $-dE/dt$ in the
strong coupling regime. To do so, we consider a $N=4$ SYM plasma instead of a QCD plasma.
The field content of this theory is 1 gauge field, 4 fermions and 6 scalars, all in the 
adjoint representation of the gauge group. Using the AdS/CFT correspondence, the quantum
dynamics of this theory at strong coupling can be obtained by classical gravity calculations
\cite{Maldacena:1997re,Witten:1998qj,Gubser:1998bc}.

\subsection{The AdS/CFT correspondence}

We consider the large $N_c,$ small gauge coupling $g_{YM}$ limit
\be
N_c\to\infty\ ,\hspace{0.5cm} g_{YM}\to0\ ,\hspace{0.5cm} \lambda\equiv g_{YM}^2 N_c\mbox{ finite,}
\ee
where the 't Hooft coupling $\lambda$ controls the theory. Then strong couping means $\lambda>>1.$
In this regime the equivalent string theory in $AdS_5$ space is weakly coupled and weakly curved:
\be
g_{YM}\ll1\Leftrightarrow g_s\ll1\mbox{ and }\lambda\gg1\Leftrightarrow R\gg l_s\ ,
\ee
where $g_s$ is the string coupling, $l_s$ is the string length and $R$ is the curvature radius of the AdS space. In this limit of small string coupling and large curvature radius, classical gravity is a good approximation of the string theory.

The background metric corresponding to the SYM theory at finite temperature is
\be
ds^2=R^2\left[\f{du^2}{h(u)}-h(u)dt+u^2(dx^i)^2\right]\equiv G_{\mu\nu}dX^\mu dX^\nu
\ee
with $h(u)=u^2-u_h^4/u^2$ and where $u_h=\pi T$ is a black-hole horizon. The corresponding Hawking temperature $T$ is the temperature of the SYM theory. The coordinate in the fifth dimension
$u=r/R^2$ has the dimension of momentum and the SYM theory lives on the boundary at
$u=\infty.$

The heavy quark, in the fundamental representation, whose energy loss we want to compute, leaves on a brane at $u=u_m=2\pi M/\sqrt{\lambda}\to\infty$ with a string attached to it, hanging down to the horizon. Points on the string can be identified to quantum fluctuations in the heavy quark wave function with virtuality $\sim u.$ Indeed, the quantum dynamics in the SYM theory is mapped onto classical dynamics in the 5th dimension. More precisely, the string dynamics is given by the Nambu-Goto action
\be
S=-\f{\sqrt{\lambda}}{2\pi R^2}\int d\tau d\sigma \sqrt{-\det g_{ab}}
\ee
where $\tau$ and $\sigma$ are the worldsheet coordinates and
\be
g_{ab}=G_{\mu\nu}(\partial_a X^\mu)(\partial_b X^\nu)
\ee
is the metric induced on the worldsheet.

\subsection{The trailing string picture}

Let us assume that the quark moves along the $x$ direction and parametrize the space-time
coordinates $X^\mu=(t,x,y,z,u)=(\tau,x(\tau,\sigma),0,0,\sigma).$ Then
\be
S=-\f{\sqrt{\lambda}}{2\pi}\int dtdu\sqrt{1-\f{u^2\dot{x}^2}{h(u)}+u^2h(u)x'^2}
\ee
with $\dot{x}=\partial_t x(t,u)$ and $x'=\partial_u x(t,u).$ The classical equation of
motion $\partial_a\delta{\cal L}/\delta\partial_a X^1=0$ gives 
\be
\f{\partial}{\partial u}\lr{\f{u^2h(u)x'}{\sqrt{-g}}}
-\f{u^2}{h(u)}\f{\partial}{\partial t}\lr{\f{\dot{x}}{\sqrt{-g}}}=0
\ee
From the solution of this equation, one gets the rate at which the energy flows
down the string:
\be
-\f{dE}{dt}=\f{\delta{\cal L}}{\delta\partial_t X^1}
=\f{\sqrt{\lambda}\ R^2}{\sqrt{-g}}u^2h(u)\dot{x}x'
\ee
which is identified to the heavy-quark energy loss.

\begin{figure}
\begin{center}
\resizebox{0.7\columnwidth}{!}{\includegraphics{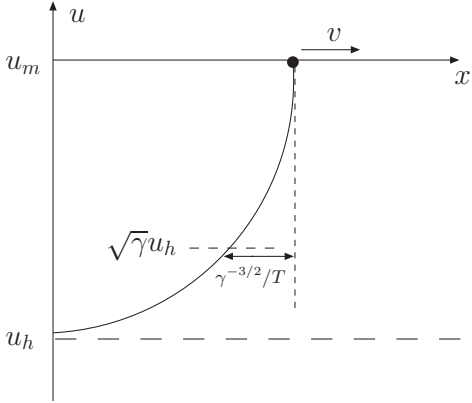}}
\caption{The trailing string solution $x(t,u)\!=\!x_0\!+\!vt\!+\!F(u).$ The part of the string
above $u\!=\!\sqrt{\gamma}u_h$ is genuinely part of the heavy quark and the part of the
string below $\sqrt{\gamma}u_h$ is emitted radiation. At $u\!=\!\sqrt{\gamma}u_h,$ the overlap
in the longitudinal direction between the trailing string and a straight string is
$\g^{-3/2} u_h.$}
\label{fig:2}
\end{center}
\end{figure}

In Ref. \cite{Herzog:2006gh,Gubser:2006bz}, the authors imagined using an external force
to pull the quark at a constant velocity $v.$
Writting $x(t,u)=x_0+vt+F(u)$ they obtained (see Fig.\ref{fig:2})
\be
F(u)=\f{1}{2u_h}\left[\f{\pi}{2}-\tan^{-1}\lr{\f{u}{u_h}}-\cot^{-1}\lr{\f{u}{u_h}}\right]
\ee
and the corresponding rate of energy loss
\be
-\f{dE}{dt}=\f{\sqrt{\lambda}}{2\pi}u_h^2 \gamma v^2\ .
\label{rate}\ee

\subsection{Introducing the saturation scale}

Investigating this picture in more details, one finds that there is a special point
on the string, at $u=\sqrt{\gamma}u_h$ \cite{Dominguez:2008vd}. In terms of quantum
fluctuations in the heavy quark wave function, this is a special momentum scale that we shall denote $Q_s$ for reasons we explain now.

First, the part of the string below $\sqrt{\gamma}u_h$ is not causally connected with the part of
the string above: this point corresponds to a black hole horizon in the rest frame of
the string. Points on the string with $u<\sqrt{\gamma}u_h$ do not know about the heavy quark. Second, the energy density around the quark is unchanged up to distances $1/(\sqrt{\gamma}u_h):$ fluctuations with a virtuality higher than $\sqrt{\gamma}u_h,$ localized within this circle, do not feel the plasma. In the fifth dimension, this means that points on the string with
$u>\sqrt{\gamma}u_h$ do not know about the plasma.

Therefore, the part of the string above $\sqrt{\gamma}u_h$ corresponds to highly virtual fluctuations still part of the heavy quark wave function while the part of the string
below $\sqrt{\gamma}u_h$ corresponds to longer-lived fluctuations which became
emitted radiation. By analogy with the weak coupling picture, one is led to call this
momentum scale the saturation scale
\be
Q_s=\sqrt{\gamma} u_h\ .
\ee
Note that with this understanding, it is straightforward to interpret the limiting velocity
phenomenon $M>\sqrt{\lambda\g}\ T.$ The maximum value of $\g$ is reached when $Q_s$ crosses the
heavy-quark brane $u=u_m.$

We now understand that quantum fluctuations in the heavy quark wave function become emitted radiation if $k_\perp<Q_s.$ The maximum energy $\omega$ a fluctuation can have can be estimated from the minimum longitudinal extend its need to overlap with the heavy quark.
At $u=Q_s,$ which is where the highest $\omega$ will be reached, the overlap in the longitudinal direction between the trailing string and a straight string (which would be the solution in the vacuum) is $1/\g Q_s$ which means $\omega<\g Q_s.$ These conditions on $k_\perp$ and $\omega$ are the same than in the pQCD case when written in terms of $Q_s,$ therefore it is no surprise that
the energy loss at strong coupling can be written (with $v=1$ at high energies)
\be
-\f{dE}{dt}\propto\sqrt{\lambda}\ Q_s^2\ .
\ee
The only differences are that the probability for the fluctuation is now $\sqrt{\lambda}$ instead of
$\alpha_s N_c,$ and of course $Q_s$ is different. This picture does not allow to determine prefactors, but it gives correct parametric dependences.

Finally, one obtains the following relation between the coherence time of the dominant radiated fluctuations $t_c=\g/Q_s$ and the saturation scale $Q_s:$
\be
Q_s=t_c T^2\ .
\ee
This is in agreement with the scale found in \cite{Hatta:2007cs} (see also \cite{Albacete:2008ze}) which separates weak from strong scattering in the deep inelastic structure functions calculations. Interestingly enough,
$1/Q_s$ is also the screening length found in \cite{Liu:2006he} which determines whether a quarkonium in the strongly-coupled SYM plasma is bounded ($d<1/Q_s$) or has dissociated ($d>1/Q_s$), with $d$ the quark-antiquark separation.

\subsection{The case of finite-extend matter}

Addressing the case of finite-extend matter is important for phenomenology. An exact calculation
is challenging, because if would require to extend the AdS/CFT correspondence to introduce the plasma length $L$ in the metric. Instead, using the picture developed above, one can infer the result. The discussion is similar to the one in pQCD: if $L>t_c=\sqrt{\g}/T,$ the results are that of the infinite matter case and if $L<t_c,$ the energy of the fluctuations which dominate the energy loss is $\omega=LQ_s^2$ and one has 
\be
-\f{dE}{dt}\propto\sqrt{\lambda}\ Q_s^2
\label{eloss}
\ee
with
\be
Q_s=L T^2\ .
\label{Qs}
\ee
Let us describe a calculation which further motivates these results \cite{Dominguez:2008vd}. A brief constant acceleration $a$ to the desired speed $v$ mimics the creation of a bare quark-antiquark pair. The accelerating string solution \cite{Mikhailov:2003er} reveals that $u=a$ is a black hole horizon \cite{Xiao:2008nr}, similarly to the Unruh effect in general relativity. Highly virtual fluctuations corresponding to $u>a$ are part of the heavy quark while the ones corresponding to $u<a$ are radiated due to the acceleration. This insures that if $a>T,$ which we assume, the medium is not felt during the creation process.

Then stopping the acceleration triggers the building of the wavefunction, essentially the separation at $u=a$ decreases as $\g/t.$ The bare quark is turning into a dressed quark while interacting with the medium. The key issue is to understand the time it takes for the heavy quark to build the fluctuations which will be freed and those that dominate the energy loss. While in pQCD this was easily evaluated because of the picture of multiple scattering locally giving transverse momentum to the radiated gluons, this is more subtle at strong coupling. The findings in \cite{Dominguez:2008vd} support the results (\ref{eloss}-\ref{Qs}). Once again this does not allow to determine prefactors, but the $L$ and $T$ dependences are robust.

\subsection{Results for $p_\perp-$broadening}

Let us give our related results for the $p_\perp-$broadening of the heavy quark
at strong coupling. Our picture gives
\be
\f{dp_\perp^2}{dt}\propto \sqrt{\lambda}\f{dQ_s^2}{dt}
\ee
where $t=t_c$ in the infinite matter case and $t=L$ in the finite matter case. Note that
because $\sqrt{\lambda}\gg1,$ radiative $p_\perp-$broadening is dominant at strong coupling, as opposed to weak coupling. Also contrary to the pQCD case, the results are different for infinite and finite matter. In the former case, one obtains
\be
\f{dp_\perp^2}{dt}\propto \sqrt{\lambda\g}\ T^3\ .
\ee
Contrary to the energy-loss formula (\ref{rate}), this result is non trivial to get with a direct calculation \cite{Gubser:2006nz,CasalderreySolana:2007qw}, but the $\gamma$ and $T$ dependence come in a straightforward manner within our picture. In the finite matter case one gets
\be
\f{dp_\perp^2}{dt}\propto \sqrt{\lambda}T^4 L\ .
\ee
Infinite and finite matter results are different at strong coupling because $p_T$ broadening is not a local phenonemon like in pQCD: there is no picture of local scatterings. At strong coupling, the transport coefficient $\hat{q}$ is not relevant, it does not control the energy loss or
$p_\perp-$broadening.

\section{Conclusions}

Heavy-quark energy loss and $p_\perp-$broadening have identical parametric form when propagating through a weakly-coupled QCD plasma or a strongly-coupled SYM plasma, when written in terms of the saturation momentum $Q_s:$
\be
-\f{dE}{dt}\propto\lr{\begin{array}{cc}\alpha_s N_c\\ \sqrt{\lambda} \end{array}}\ Q_s^2
\hspace{0.5cm}
\f{dp_\perp^2}{dt}\propto\lr{\begin{array}{cc}\alpha_s N_c\\ \sqrt{\lambda} \end{array}}\ 
\f{dQ_s^2}{dt}
\hspace{0.5cm}
\begin{array}{cc} QCD \\ SYM \end{array}\ .
\ee
This is not surprising as $Q_s,$ in both cases, is defined as the scale with controls which heavy quark fluctuations become emitted radiation. While for a pQCD plasma it is known that such a scale can be singled out, the fact that one could identify one also at strong coupling is the main message of this work. Then the different couplings $\alpha_s N_c$ and $\sqrt{\lambda}$ simply reflect the fluctuation probabilities in the heavy quark wave function. The saturation scale $Q_s$ is given by
\begin{itemize}
\item for infinite-extend matter or $L>t_c=\g/Q_s$
\be
Q^2_s=(\hat{q}\gamma)^{2/3}\mbox{ in pQCD and }
Q^2_s=T^2\gamma\mbox{ in SYM}\ ,
\ee
\item for finite-extend matter with $L<t_c$
\be
Q^2_s=\hat{q}L\mbox{ in pQCD and }
Q^2_s=T^4 L^2\mbox{ in SYM}\ .
\ee
\end{itemize}
We recall that $\hat{q}\sim\alpha_s T^3$ in pQCD.

On the right-hand side on the $p_\perp-$broadening formula, $t=t_c$ in the infinite matter case and $t=L$ in the finite matter case. The pQCD result $dp_\perp^2/dt=\alpha_s N_c\hat{q}$ is for radiative $p_\perp-$broadening and is actually subdominant ($\alpha_s\ll1$), $p_\perp-$broadening is multiple scattering dominated in pQCD $dp_\perp^2/dt=\hat{q}.$ In SYM, because
$\lambda\gg1,$ $p_\perp-$broadening is radiation dominated. Hence this is a non local phenomenon and the transport coefficient $\hat{q}$ is not a relevant quantity at strong coupling. However if in the pQCD energy loss problem the soft scales in the process are subject to a strong effective coupling, $\hat{q}$ will be enhanced by that dynamics \cite{Mueller:2008zt}.

Finally, we obtained the first estimate of the plasma length dependence at strong coupling
of the heavy-quark energy loss and $p_\perp-$broadening. The conclusion is that the $L$ dependence in much stronger at strong coupling, ($\Delta E\sim L^3,\Delta p_\perp^2\sim L^2$) in SYM
while ($\Delta E\sim L^2,\Delta p_\perp^2\sim L$) in pQCD. We also note the results discussed in \cite{Kharzeev:2008qr}, that the energy loss at strong coupling will behave as $\Delta E\sim L,$ if the relevant dynamics is that of early times, before thermalization, while in this work we considered a thermalized medium.

%

\end{document}